\documentclass{aa}
\usepackage[dvips]{graphics}
\usepackage{latexsym}
\usepackage{graphicx}
\usepackage{subfigure}

\def\msun{\hbox{$M_\odot$}}
\def\gtsima{$\; \buildrel > \over \sim \;$}
\def\ltsima{$\; \buildrel < \over \sim \;$}
\def\prosima{$\; \buildrel \propto \over \sim \;$}
\def\gsim{\lower.5ex\hbox{\gtsima}}
\def\lsim{\lower.5ex\hbox{\ltsima}}
\def\simgt{\lower.5ex\hbox{\gtsima}}
\def\simlt{\lower.5ex\hbox{\ltsima}}
\def\simpr{\lower.5ex\hbox{\prosima}}

\begin{document}

\thesaurus{03(11.05.1; 11.05.2; 11.06.1)}

\title{EROs and the formation epoch of field ellipticals}

\author{
	E. Daddi \inst{1} 
	\and
	A. Cimatti \inst{2}
	\and
	A. Renzini \inst{3}
}
\institute{ 
	Universit\`a degli Studi di Firenze, 
	Dipartimento di Astronomia e Scienza dello Spazio,
	Largo E. Fermi 5, I-50125 Firenze, Italy
 	\and  
	Osservatorio Astrofisico di Arcetri,
	Largo E. Fermi 5, I-50125 Firenze, Italy
	\and
	European Southern Observatory, D-85748 Garching, Germany
} 

\offprints{edaddi@arcetri.astro.it}
\date{Received ; accepted }

\maketitle
\markboth{Daddi, Cimatti \& Renzini}{EROs and the formation epoch of
field ellipticals}

\begin{abstract}

A comparison is presented between the observed surface density
of extremely red objects (EROs) and the densities of red 
high-$z$ elliptical galaxies predicted by Pure Luminosity 
Evolution (PLE) and by some CDM hierarchical models. The analysis 
is based on  the widest field survey for EROs available to 
date (covering in total $\sim$850 arcmin$^2$), and it takes into account
the effects of ERO clustering. Good agreement is found between
the observed surface densities of EROs and those predicted by
PLE models for ellipticals selected using the same color and luminosity
thresholds.
Since there is evidence that the bulk of EROs are passively evolving 
ellipticals, this result implies that most field ellipticals were fully
assembled at least by $z\sim 2.5$. Existing hierarchical models predict
instead
a surface density of red ellipticals that is much smaller than that of
the EROs. 

\end{abstract}

\section{Introduction}

The formation of elliptical galaxies remains one of the most
controversial issues of galaxy evolution and structure
formation. While there is now wide consensus that the bulk of stars in
ellipticals formed at very high redshifts (see e.g. Renzini 1998 and the introduction of
Schade et al. 1999 for recent reviews), opinions diverge as to the epoch
at which ellipticals have been assembled to their present size and mass.  
In semianalytical simulations of hierarchical Cold Dark Matter (CDM)
cosmologies (e.g. Kauffmann 1996; Baugh, Cole \& Frenk 1996;
Baugh et al. 1998) massive ellipticals form at
relatively low redshift ($z<1$) through the merging of spiral galaxies.

Empirical support for this scenario has been claimed from an alleged
deficit of old ellipticals at $z\sim 1$ compared to pure luminosity evolution
(PLE) models (Kauffmann, 
Charlot \& White 1996, Zepf 1997, Franceschini et al. 1998, 
Barger et al. 1999, Menanteau et al. 1999, Treu \& Stiavelli 1999).
However, others found evidence for a constant comoving density of 
ellipticals up to $z\sim 2$ (Totani \& Yoshii 1997, Benitez et al. 1999, 
Broadhurst \& Bowens 2000, Schade et al. 1999; Scodeggio \& Silva 2000),
which would instead favor an early assembly of ellipticals, no matter
whether in clusters or in the field. 
One obvious way to solve the issue is to search for putative high-$z$ ellipticals
over much larger fields than in previous studies. 
Passively evolving ellipticals in the redshift range $1\simlt
z\simlt 2$ are in fact characterized by very red colors ($R-K_{\rm s}
\gsim 5-6$), which qualify them as extremely red objects (ERO). 

In a recent study based on the largest survey for EROs to date, Daddi et al. (2000) reported
the discovery of strong EROs angular clustering.
This finding suggest that the origin of such discrepant results were definitively in
the small fields covered by previous studies (ranging from $\sim 1$ to $\sim60$
arcmin$^2$), which are likely to be affected by 
field-to-field variations in the number of intercepted large scale
structures, hence in the number of red, color selected galaxies. 
Moreover, the strong clustering argues
for the bulk of EROs being made by elliptical (Daddi et al. 2000)
rather than starburst galaxies, that can have very red $R-K_{\rm s}$ colors as well 
(e.g. Cimatti et al. 1998). This agrees with independent indications that
ellipticals are at least a fraction of $\sim70$\% among EROs (Moriondo et al. 2000,
Stiavelli et al. 2000 private comunication, Cimatti et al. 1999).
Under this assumption, the surface density of EROs can therefore set 
strong constraints to models for the formation and evolution of elliptical galaxies.

In this Letter we present the results of the comparison
between the observed surface density of EROs and the densities
predicted by PLE and hierarchical models for the evolution of the elliptical
galaxy population. Such a comparison is based on the results of our wide-field survey 
for EROs (Daddi et al. 2000) and on the Thompson et al. (1999) survey, covering a total 
area of about 850 arcmin$^2$ resulting in the selection of a total sample 
of several hundreds of EROs. 
In addition, the measurements of the angular clustering allow us to quantify for the first time 
the uncertainties on the surface density of EROs due to cosmic variance. 
For these reasons, our study represents a significant improvement respect 
to previous attempts. For example, the widest field analysis most
recently done was based only on 16 EROs selected from an area of 
60 arcmin$^2$ (Barger et al. 1999). 

\section{Predicting the surface density of high-$z$ ellipticals}

The surface density of EROs ($\mu$, in objects arcmin$^{-2}$) redder that a
color threshold $T$, in a survey with limiting magnitude $K_{\rm LIM}$
is calculated as:
\begin{displaymath}
\mu = \int_0^\infty \!\! dz_{\rm f}\int_{z_{\rm a}(z_{\rm f}, T)}^{z_{\rm b}(z_{\rm f}, T)}\!\!\!\!\!\!
dz \int_{-\infty}^
{M_{\rm UP}} \hspace{-.3truecm} dM\; LF(M,z,z_{\rm f}) \frac{dV(z)}{dz}F(z_{\rm f})
\end{displaymath}
where $F(z_{\rm f})$ describes the formation redshift distribution 
(for a single redshift of formation $z_{\rm f}=z_0$, $F(z_{\rm f})=\delta(z_{\rm f}-z_0)$),
$z_{\rm a}$ and $z_{\rm b}$ give the redshift range
where the galaxy is redder than the threshold $T$ (derived from the
modeled $R-K_{\rm s}$ color vs. $z$), $M_{\rm UP}$ is the
absolute $K_{\rm s}$ magnitude of the faintest galaxy detectable in the 
survey at a given $z$ (calculated from $K_{\rm LIM}$ via the luminosity 
distance and the K--correction), $V(z)$ is the comoving volume 
intercepted by 1 arcmin$^2$ and LF is the luminosity function where 
$M^{\star}$ evolves consistently with $z$. 

The following ingredients  are required in order to compute $\mu$
in a PLE scenario: 
a spectral synthesis model, an IMF, a metallicity, a star formation 
history (i.e. how the initial burst is characterized and how the star
formation rate evolves with time), a formation redshift $z_{\rm f}$,
a local LF for elliptical galaxies, a dust reddening and, finally, a 
set of cosmological parameters. Our calculations are based on the 
Bruzual \& Charlot (1997) spectral synthesis models with the Salpeter IMF, 
solar metallicity and no dust reddening. 
Following Barger et al. (1999), the Marzke et al. (1994) pure elliptical 
LF was used, transforming the magnitudes from the $B$ to the $K$-band assuming
$B-K=4.43$ as the color of local ellipticals (Huang et 
al. 1998). 
The results discussed in the next sections do not
significantly change by using instead a  $K$--band LF 
(e.g., Loveday 2000; Szokoly et al. 1998; Gardner et al. 1997;
Glazebrook et al. 1995; Mobasher et al. 1993), or the $R$--band LF 
(Lin et al. 1996) adopting $R-K\sim2.7$ for local ellipticals and 
a fraction of ellipticals in those samples of 25--30\% of the 
total population of galaxies. 
We adopt $\Omega_0=0.3, \Lambda=0.7$, and $h_{100}=0.7$, for which the present age of the universe is 
$\sim14$ Gyr. A wide range of formation redshifts was  considered, 
from $z_{\rm f}=1$ to $z_{\rm f}=30$. The star formation rate is assumed to decline
exponentially following the beginning of a burst at $z=z_{\rm f}$, 
with a timescale $\tau=0.1$ and 0.3 Gyr. 

\section{Comparison with the observations}

The database consists of the Daddi et al. (2000) and of the Thompson et al. 
(1999) surveys. The first one is the widest survey for EROs so far and is made of a single
mosaic
centered at $\alpha=14^{\rm h}49^{\rm m}29^{\rm s}$ and 
$\delta=09^o00^{\prime}00^{\prime\prime}$ (J2000); it
is based on $K_{\rm s}$ and $R$
band imaging and it is complete to $K_{\rm s}=18.8$ over the whole area of 701 arcmin$^2$ and to
$K_{\rm s}=19.2$ over 447.5 arcmin$^2$. 
The latter survey is smaller (154 arcmin$^2$ centered at 
$\alpha=16^{\rm h}24^{\rm m}33^{\rm s}$ and
$\delta=55^o43^{\prime}59^{\prime\prime}$ (J2000)) but, being complete to $K^\prime\sim20$,
it has been used to extend the magnitude range to deeper values,  providing sufficient
statistics. 

EROs were selected using the color thresholds $R-Ks>5.3$ and $R-Ks>6$, 
which correspond to passively evolving ellipticals at $z\simgt1$ and 
$z\simgt1.3$, respectively (Daddi et al. 2000). An $L=L_*$ elliptical evolved at
$z=1$ and $z=1.3$ would have $K_{\rm s}\sim18.5$ and $K_{\rm s}\sim19.2$, respectively, thus our
samples consist of bright $L\approx L_*$ objects.
Fig. 1 shows the resulting surface densities of EROs. 
After taking into account the different filters used, the Thompson et al. survey yields a 
surface density of EROs of 0.42 and 0.8 arcmin$^{-2}$ 
for the $R-K_{\rm s}>5.3$ samples with $K_{\rm s}<19.2$ and $K_{\rm s}<19.6$ respectively, and 
0.4 arcmin$^{-2}$ for the $R-K_{\rm s}>6$ sample with $K_{\rm s}<19.8$. 

The use of such large surveys is of crucial importance to carry out a reliable comparison.
In fact Daddi et al. (2000) showed that, as EROs are strongly clustered and large voids are present
in their distribution on the sky, small field surveys can easily lead to wrong determinations of
the ERO surface density and the probability of an underestimate is much larger than that of an
overestimate. As an example of this effect, in the small (43 arcmin$^2$) Chandra/AXAF 
Deep Field South the surface density of EROs with $R-K_{\rm s}>5$ and $K<19$,
derived by Scodeggio \& Silva (2000), is  a factor of 5 smaller than the one derived in the
700 arcmin$^2$ survey by Daddi et al. (2000), with the same color and magnitudes thresholds.

\begin{figure*}[hp]
\label{fig:compar}
\caption{\footnotesize 
The three panels show the observed surface 
densities of EROs as derived from Daddi et al. (2000) and Thompson 
et al. (1999) surveys (filled circles and empty squares respectively).
The curves show the predictions of the surface density of ellipticals
with colors redder than $R-K_{\rm s}=5.3$ (left) and than $R-K_{\rm s}=6$ (right). 
Panel (a): case of passively evolving stellar populations
formed with \mbox{$\tau=0.1$} Gyr, at $z_{\rm f}=1.9, 2, 2.5, 3, 10$. 
Panel (b): case of passively evolving stellar populations
formed with \mbox{$\tau=0.3$} Gyr, at $z_{\rm f}=2.4, 2.5, 3, 3.5, 10$. Panel (c): 
Predictions in the case ellipticals form at $2<z_{\rm f}<3$ (for $\tau=0.1$ Gyr) and
at $2.7<z_{\rm f}<3.7$ (for $\tau=0.3$ Gyr). The comparison of the observations 
with the models is described in the text.
}
    \subfigure{
    \label{mini1}
    \begin{minipage}[b]{0.5\textwidth}
        \centering \includegraphics[height=15cm,angle=-90]{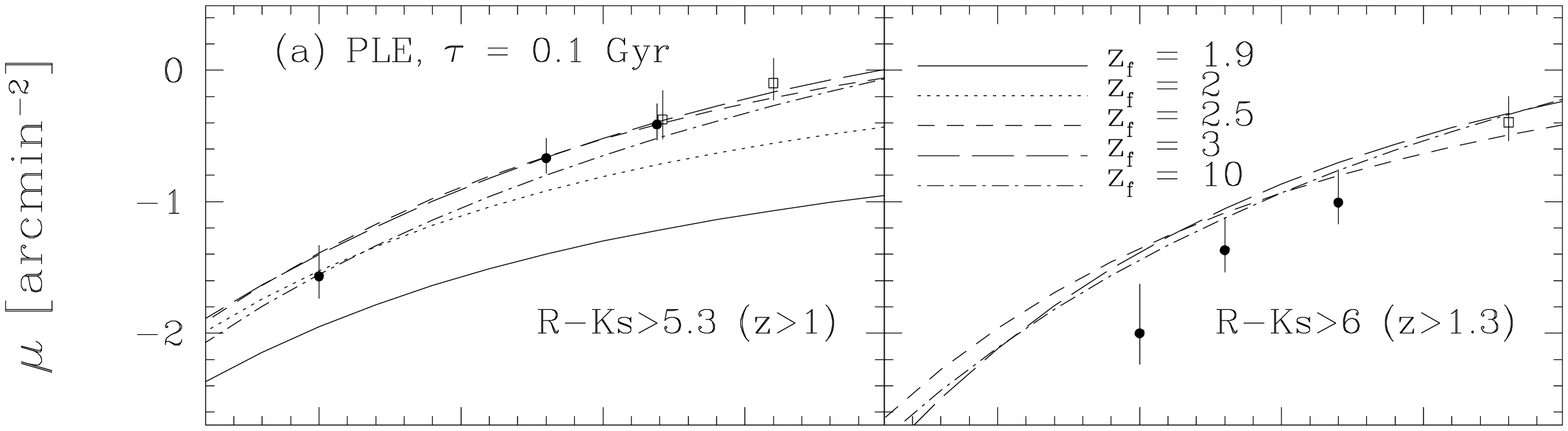}
    \end{minipage}}\\

    \vspace{-1.025truecm}
    \subfigure{
    \label{mini2}
    \begin{minipage}[b]{0.5\textwidth}
        \centering \includegraphics[height=15cm,angle=-90]{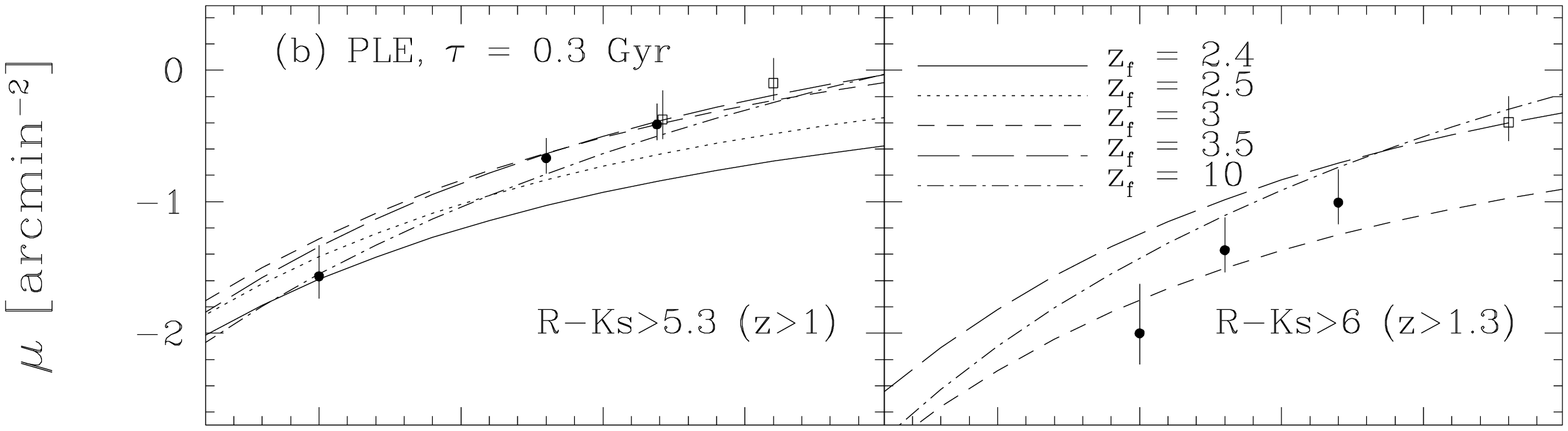}\\
    \end{minipage}}\\

    \vspace{-1.025truecm}
    \subfigure{
    \label{mini3}
    \begin{minipage}[b]{0.5\textwidth}
        \centering \includegraphics[height=15cm,angle=-90]{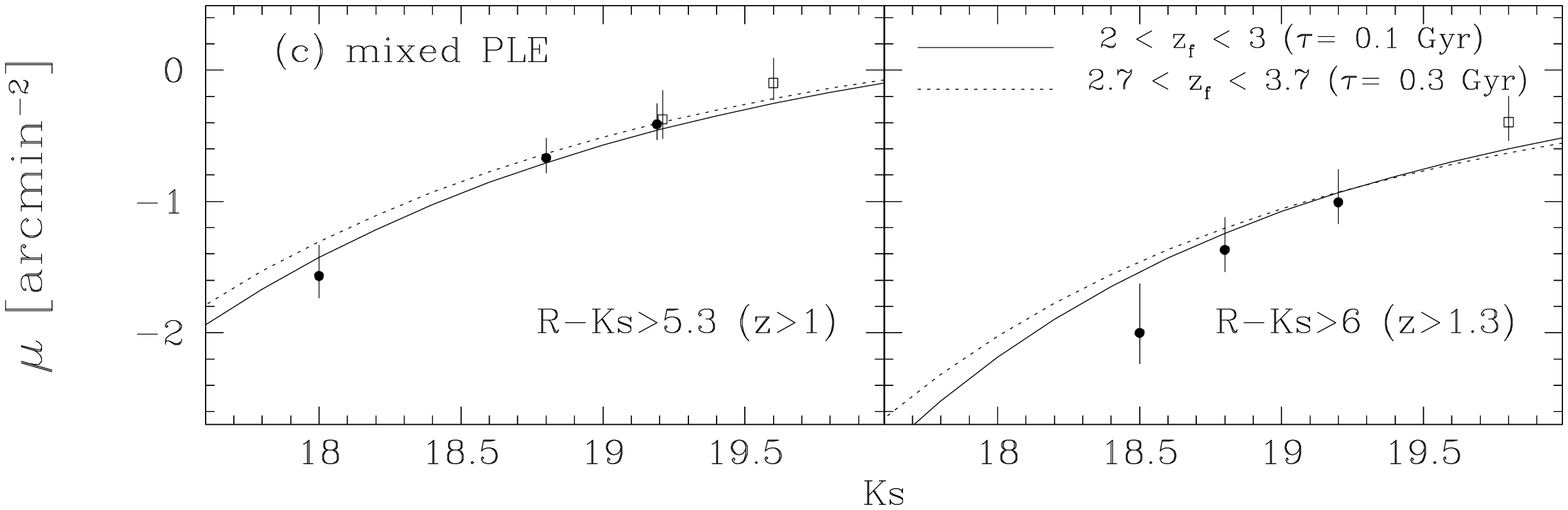}
    \end{minipage}}
\end{figure*}

Given the clustering effect, error bars in the surface density of EROs are
larger than the mere Poisson statistics for the number of objects.
The extreme of error bars shown in Fig. 1 are obtained
by solving for $\overline{n}$ the equation
$(\overline{n}-N)^2=\overline{n}(1+AC\overline{n})$, where $\overline{n}$ and $N$
are respectively the expected average and the observed number of 
ellipticals in the field, $A$ is the clustering amplitude and $C$
is a factor depending on the geometry (see Daddi et al. 2000 for more
details). A direct measure of $A$ for the EROs
with $R-K_{\rm s}>6$ is not available given the relatively small number of objects in
the survey. We therefore used an extrapolation of the relation between $A$ 
and $R-K_{\rm s}$ obtained by Daddi et al. (2000) for $3.0\leq R-K_{\rm s}\leq5.7$
and $K_{\rm s}<19.2$.
In order to estimate the clustering amplitude for the EROs with $K_{\rm s}>19.2$ 
taken from the Thompson et al. survey, as it was suggested that
$A$ decreases with $K_{\rm s}$ (Daddi et al. 2000), we conservatively
used the $K_{\rm s}\sim 19$ value, lowered by a factor of 2. 
Fig. 1 (panels a-b) shows the comparison between
the PLE model predictions and the observed surface density of EROs.

$\bullet$ {\it $R-K_{\rm s}>5.3$}: this color threshold corresponds to the
selection of $z>1$ passively evolving ellipticals, and the plots
show good agreement, for the models with high formation redshift, 
between observed and predicted surface
densities, i.e., the density of EROs is consistent
with the  expectation of PLE of local ellipticals.  Models in which
ellipticals formed at $z_{\rm f}<2.5$ ($\tau=0.1$ Gyr) or at $z_{\rm f}<3$
($\tau=0.3$ Gyr) are ruled out because they strongly underpredict the
density of EROs.
The results do not significantly change if one
assumes that a fraction up to 30\% of the EROs with $R-K_{\rm s}>5.3$ are
dusty starburst galaxies, low
mass stars, or brown dwarfs (the observed densities shown in Fig. 1
would decrease only by 0.15 dex). 

$\bullet$ {\it $R-K_{\rm s}>6$}: this color threshold corresponds to 
$z>1.3$ passively evolving ellipticals. Only the models with $z_{\rm f}>2.5$ 
($\tau=0.1$ Gyr) or  with $z_{\rm f}>3$ ($\tau=0.3$ Gyr) succeed in reproducing the observed
surface
densities of EROs, also for this redder threshold.
Fig. 1 shows that the densities predicted by the models with the highest $z_{\rm f}$ are 
generally higher than the observed ones, but only at $\sim 1\sigma$ 
significance level. Our data do not
allow us to conclude whether this is a real deficit of the reddest
ellipticals, or if it is due to the poorer statistics with respect
to the $R-K_{\rm s}>5.3$ sample (typically $\sim40$ objects with $R-K_{\rm s}>6$ 
vs. $\sim200$ with $R-K_{\rm s}>5.3$). 

Assuming that this deficit of $R-K_{\rm s}>6$ ellipticals is real, this would
imply that the formation redshift must be close to $z_{\rm f} \sim 2.5$
(for $\tau=0.1$ Gyr) or $z_{\rm f} \sim 3$ (for $\tau=0.3$ Gyr). 
To check this point we also 
tested a model where ellipticals formed with a constant rate per 
unit $z$ in the range of $2<z<3$ ($\tau=0.1$ Gyr), or $2.7<z<3.7$ 
($\tau=0.3$ Gyr). 
Fig. 1 (panel c) shows that this ``mixed PLE'' scenario
provides a good agreement with the observations of the $R-K_{\rm s}>6$ EROs.

These results 
do not significantly change in case 
of an open universe ($\Omega=0.1$, $\Lambda=0$,  $h_{100}=0.7$) or for a flat universe ($\Omega=1$, $\Lambda=0$, $h_{100}=0.7$).

Existing semiempirical models of hierarchical galaxy formation imply a
specific surface density of EROs to various limiting magnitudes.
However, no such numbers have been explicitly calculated so far, which
makes less straightforward the comparison with the observed number of
EROs. According to Kauffmann et al. (1999, Fig. 9 and 10; see also Kauffmann \& Charlot
1998b), 
in their $\tau$CDM model 
the number of massive ellipticals (stellar mass greater than $10^{11}\msun$)
decreases by a factor of 3 by $z=1$ and by a factor of $\sim 15$ by $z=2$.
Such massive ellipticals are brighter than $K=19$ for $1<z<1.5$
(Kauffmann \& Charlot 1998a). Hence this specific realization greatly 
underpredicts the numbers of EROs compared to the empirical values.
The comparison is less clear-cut in the case of  the $\Lambda$CDM 
model model of Kauffmann et al. (1999), where the number of massive
ellipticals decreases by only $\sim 30\%$ by $z=1$ and by a factor of $\sim 3$
by $z=2$. However, only a fraction of such {\it ellipticals} would qualify as
EROs, given that in the adopted nomenclature they form by merging gas-rich
spirals, with the gas being instantly converted into stars. Therefore, most 
such $1<z<1.5$ `ellipticals'  contain a significant fraction of young
stars, likely to make them bluer than the color threshold for EROs.
\section{Conclusions}

A robust surface density of EROs has been obtained from a field that
is at least ten times larger than in previous determinations, thus
greatly reducing the influence of the strong field-to-field variations
(cosmic variance), that we estimate given the known clustering
properties of EROs (Daddi et al. 2000).  This observed surface density
of EROs is in fine agreement with the density of $z>1$ ellipticals with the 
same color and luminosity thresholds as predicted by a PLE model in which 
the bulk of
field ellipticals were already fully assembled at least by $z = 2.5$, and evolved 
passively thereafter. The agreement with the predictions of the PLE models 
remains good even if some 30\% of EROs are not elliptical galaxies 
(e.g. dusty starburst galaxies). This result argues for much of the merging
being already completed by $z\gsim 2$, in such a way to make elliptical
galaxy formation mimicking the old monolithic collapse models.

A more thorough comparison with current realizations of hierarchical models
of galaxy formation is hampered by the lack of published predictions
for the surface density of EROs. In these models the number density of
massive elliptical galaxies drops rapidly with redshift. It does so
catastrophically in a $\tau$CDM model (Kauffmann \& Charlot 1998b),
which therefore can be readily excluded, unless only a very small fraction
of EROs is made of passively evolving ellipticals. However, this would
be in contradiction with the reported strong clustering properties of EROs
(Daddi et al. 2000). Less dramatic is the comparison with a
$\Lambda$CDM model, in which the number of ellipticals drops more
gently with redshift. Nevertheless, we argue that also this model can
hardly account for the observed number of EROs unless merging does
{\it not} result in significant star formation, which is not the case
in existing realizations. 
Future semianalytical models 
including specific predictions concerning the number densities of EROs
would offer an interesting opportunity, as a comparison with the observed 
densities is expected to provide a stringent test for such models. 

\begin{acknowledgements}
We thank the referee, Massimo Stiavelli, for useful comments.
\end{acknowledgements}

\end{document}